\documentclass{cernyrep}
\begin{document}
\title{The LAGUNA design study-- towards giant liquid based underground
detectors for neutrino physics and astrophysics and proton decay \\
searches\footnote{Contribution to the Workshop ``European Strategy for Future Neutrino Physics'', CERN, Oct. 2009, to appear in the Proceedings.}}

\author{{\bf The LAGUNA consortium}\thanks{LAGUNA Coordinator : andre.rubbia@cern.ch}: D. Angus$^a$, A.~Ariga$^b$, D.~Autiero$^c$, A.~Apostu$^d$, A.~Badertscher$^e$, T.~Bennet$^f$, 
G.~Bertola$^g$, P.F.~Bertola$^g$, O.~Besida$^h$, A.~Bettini$^i$
% ...
}

\institute{\small (a) University of Durham (UDUR), University Office,
Old Elvet, Durham DH1 3HP, United Kingdom
(b) University of Bern, 4 Hochschulstrasse, CH-3012,
Bern 
(c) Centre National de la Recherche Scientifique,
Institut National de Physique Nucl\'eaire et de Physique des Particules
(CNRS/IN2P3), 3 rue Michel-Ange, Paris 75794, France 
(d) Horia Hulubei National Institute of RD for Physics and Nuclear
Engineering, IFIN-HH, 407 Atomistilor Street, R-077125, Magurele,
jud. ILFOV, PO Box MG-6, postal code RO-077125, Romania
 (e) ETH Zurich, 101 Raemistrasse, CH-8092 Zurich
(f) The University of Sheffield (USFD), New Spring House 231, Glossop
Road, Sheffield S102GW, United Kingdom
 (g) Lombardi Engineering Limited, via R.Simen, CH-6648, Minusio
 (h) Commissariat \`a l'Energie Atomique (CEA)/ Direction des Sciences
de la Mati\`ere, 25 rue Leblanc, Paris 75015, France
(i) Laboratorio Subterraneo de Canfranc (LSC), Plaza del Ayuntamiento
no. 1, 22880 Canfranc (Huesca), Spain
 (j) Mineral and Energy Economy Research Institute of the Polish
Academy of Sciences (IGSMIE-PAN), Wybickiego 7, 30-950 Krakow, Poland
 (k) Wroclaw University of Technology (PWr Wroclaw), ul. Wybrzeze
Wyspianskiego 27, 50-370 Wroclaw, Poland
 (l) University of Bucarest (UoB), Faculty of Physics Bld.Atomistilor
nr.405, Physics Platform, Magurele, Ilfov County, RO-077125, MG-11
Bucharest-Magurele, Romania
(m) University of Oulu (U-OULU), 1 Pentti Kaiteran Katu, Oulu 90014,
Finland
 (n) Technische Universit\"at München (TUM), 21 Arcisstrasse, M\"unchen
80333, Germany
(o) University of Aarhus (AU), 1 Norde Ringgade, Aarhus C 8000,
Denmark
 (p) AGT Ingegneria Srl, Perugia, 10 A via della Pallotta, Perugia
06126, Italy
 (q) Technodyne International Ltd., Unit16, Shakespeare Business
Centre Hathaway Close, Eastleigh UK SO 50 4SR, United Kingdom 
 (r) Kalliosuunnittelu Oy Rockplan Ltd., 2 Asemamiehenkatu, Helsinki
00520, Finland
 (s) University of Jyv\"askyl\"a (JyU), 9 Survontie, Jyv\"askyl\"a 40014,
Finland 
 (t) Cleveland Potash Limited (CPL), Boulby Mine, Loftus, Saltburn
Cleveland, TS13 4UZ, UK
 (u) Institute of Physics, University of Silesia Uniwersytecka 4, 40-007 Katowice, Poland
 (v) Universidad Autonoma de Madrid (UAM), C/Einstein no. 1; Rectorado,
Ciudad Universitaria de Cantoblanco, 28049 Madrid, Spain
(w) Max-Planck-Institute for Nuclear Physics, Heidelberg
 (x) KGHM CUPRUM Ltd Research and Development Centre, Pl. 1 Maja,
50-136 Wrocaw, Poland
 (y) IFJ Pan, H.Niewodniczaski Institute of Nuclear Physics PAN,
Radzikowskiego 152, 31-342 Krakow, Poland
(z) Max-Planck-Institute for Physics, Munich
(A) High Energy Physics Department - A. Soltan Institute for Nuclear
Studies (SINS) Hoza 69 00-681 Warsaw, Poland
 (B) Faculty of Physics and Astronomy, Wroclaw University, pl M. Borna 9, 50-204 Wroclaw, Poland.}
\maketitle

\begin{abstract}
The feasibility of a next generation neutrino observatory in Europe is
being considered within the LAGUNA design study. To
accommodate giant neutrino detectors and shield them from cosmic rays, 
a new very large underground infrastructure is required.
Seven potential candidate sites in different parts of Europe and at several distances
from CERN are being studied: Boulby (UK), Canfranc (Spain), 
Fr\'ejus (France/Italy), Pyh\"asalmi
(Finland), Polkowice-Sieroszowice (Poland), Slanic (Romania) and Umbria (Italy). 
The design study aims at the comprehensive and coordinated technical assessment of each site, 
at a coherent cost estimation, and at a prioritization of the sites within the summer 2010.
\end{abstract}

\section{Physics goals}
Large underground neutrino detectors, for instance SuperKamiokande or SNO, have achieved fundamental results
in particle and astro-particle physics.
A next-generation very large multipurpose 
neutrino observatory of a total mass in the range of 100'000 to 1'000'000 tons will 
provide new and unique scientific opportunities in this field, very likely leading to fundamental discoveries.
It will aim at a significant improvement in the sensitivity to search for proton decays, pursuing 
the only possible path to directly test physics at the GUT scale, 
extending the proton lifetime sensitivities up to 10$^{35}$ years, a range compatible 
with several theoretical models;
 it will measure with unprecedented sensitivity the last unknown mixing angle $\theta_{13}$ 
 and unveil the existence of CP violation in the leptonic sector, which in turn could provide 
 an explanation of the matter-antimatter asymmetry in the Universe; moreover it will detect 
 neutrinos as messengers from astrophysical objects as well as from the Early Universe to give us information on 
 processes happening in the Universe, which cannot be studied otherwise. 
 In particular, it will sense a large number of neutrinos emitted by
exploding galactic and extragalactic type-II supernovae, allowing an accurate study of the mechanisms driving the
explosion. The neutrino observatory will also allow precision studies of other astrophysical or terrestrial sources of
neutrinos like solar and atmospheric ones, and search for new sources of astrophysical neutrinos, like for example
the diffuse neutrino background from relic supernovae or those produced in Dark Matter (WIMP) annihilation in the
centre of the Sun or the Earth.
 
 \section{The LAGUNA design study}
The construction of a large scale neutrino detector in Europe devoted to particle and astroparticle physics was discussed
 several years ago (see e.g.~\cite{Autiero:2007zj}) and is currently one of the priorities of the ASPERA roadmap, defined in 
 2008~\cite{aspera}.  Four national underground laboratories located resp. 
 in Boulby (UK), Canfranc (Spain), Gran Sasso (Italy), and Modane (France), are today in operation, 
 hosting detectors looking for Dark Matter or neutrino-less double beta decays, or performing long-baseline experiments. 
 However, none of these existing laboratory is large enough for the next-generation neutrino experiments.

The FP7 Design Study LAGUNA ~\cite{laguna} (Large Apparatus studying Grand Unification and Neutrino Astrophysics)
  is an EC-funded project carrying on underground sites studies and developments in view of such detectors observatories.
 Three detector options are currently being studied: GLACIER~\cite{Rubbia:2004tz}, LENA~\cite{Oberauer:2005kw}, 
 and MEMPHYS~\cite{deBellefon:2006vq}.
The study is evaluating possible extensions of the existing deep underground laboratories, 
and on top of it, the creation of new laboratories in the following regions: Umbria Region (Italy), Pyh\"asalmi (Finland), 
Sieroszowice (Poland) and Slanic (Romania). 
Table ~\ref{tab:1} 
summarizes some basic characteristics of the sites under consideration, including the distance from CERN, which 
is relevant in case a neutrino beam is sent from CERN to the selected underground site.

\begin{table}[h]
\begin{center}
\caption{Potential sites being studied with the LAGUNA design study.}
\label{tab:1}
\begin{tabular}{ccccc}
\hline\hline
\textbf{Location} 	& \textbf{Type} & \textbf{Envisaged depth}	 
								& \textbf{Distance from}	
									& \textbf{Energy 1$^{st}$ Osc. Max.}	 	\\
				&	& \textbf{m.w.e.} 	& \textbf{CERN [km]}		& \textbf{[GeV]} 	\\												
\hline
Fr\'ejus (F) 		& Road tunnel  &  $\simeq$ 4800 & 130 & 0.26   \\
Canfranc (ES)  		& Road tunnel  &  $\simeq$ 2100  & 630 & 1.27  \\
Umbria(IT) $^a$ 		& Green field    &  $\simeq$ 1500  & 665  & 1.34  \\
Sieroszowice(PL)  	& Mine 	         &  $\simeq$ 2400  & 950 & 1.92  \\
Boulby (UK)	  	& Mine 		&  $\simeq$ 2800  & 1050 & 2.12  \\
Slanic(RO)	  	& Salt Mine 	&  $\simeq$ 600  & 1570 & 3.18  \\
Pyh\"asalmi (FI)	& Mine 		&  up to $\simeq$ 4000  & 2300 & 4.65  \\
\hline\hline
\multicolumn{3}{l}{$^{a}$ \footnotesize{$\simeq$1.0 $^{\circ}$ off axis.}}
\end{tabular}
\vspace*{-7mm}
\end{center}
\end{table}

\section{Site selection}

Site selection is a complex process involving the optimization and assessment of several parameters,
encompassing physics performance, technical feasibility, safety and legal aspects, socio-economic and environmental impact,
costs, etc.
As a result, LAGUNA is interdisciplinary, involving most European physicists interested in realizing
massive neutrino detectors, as well as 
 geo-technical experts, geo-physicists, structural engineers, mining engineers and also large storage tank engineers.
It  regroups 21 beneficiaries, composed of academic institutions from Denmark, Finland, France, Germany, Poland,
Spain, Switzerland, United Kingdom, as well as industrial partners specialized in civil and mechanical engineering
and rock mechanics, commonly assessing the feasibility of a this Research Infrastructure in Europe.
Some of the typical issues addressed by LAGUNA are illustrated below:
\begin{enumerate}
\item assessment of strengths and weaknesses  of each site;
\item rock mechanics and excavation of the required caverns;
\item overburden vs. detector options;
\item design and construction of tanks in relation to sites;
\item  transport, access, delivery of detector liquids;
\item safety issues, in road tunnels and in operating mines;
\item environmental impact, e.g. rock removal;
\item relative costs;
\item etc.
\end{enumerate}

The study also gives the opportunity to the members of the consortium to visit all the potential
sites and hold several meetings to exchange ideas and work together on common areas, solving 
common issues, in a collaborative spirit, although a sense of healthy competition is implied
among the sites.

The results will be summarized in 16 reports (``deliverables") to be submitted to the EC within the fall 2010.
A first report on the health, safety and environmental impact has been prepared during the summer 2009.
Seven ``interim'' reports, one for each site, are currently under preparation and should be available
during the winter 2009/2010. They demonstrate
that the site studies are well underway with definitions of the
infrastructure in advanced stages, providing well-defined conceptual designs 
and reliable excavation cost-estimates.

\section{The future of long baseline neutrino physics in Europe}

The next generation deep underground neutrino detector
should be coupled to advanced neutrino beams from CERN
to complete the understanding of the leptonic mixing matrix, in particular
to study matter-antimatter asymmetry in neutrino
oscillations (CP-violation), thereby addressing the outstanding puzzle of the origin of the excess of matter over 
antimatter created in the very early stages of evolution of the Universe. The outcome of the current international program 
to measure the (small) mixing angle $\theta _{13}$ (T2K in Japan, DoubleChooz in Europe, Daya Bay in China 
and NOvA in the US) will define the strategy to measure CP-violation in the neutrino sector, whether using conventional 
superbeams or more exotic beta-beams or neutrino factories. 
In the case that $\theta_{13}$ is below the sensitivity of T2K/DoubleChooz/Daya Bay/NOvA, 
a new far detector designed to measure CP-violation coupled to
an intensity upgraded conventional superbeam, is 
the most natural next step to continue exploring neutrino oscillations, since the increased statistics
yields an order of magnitude better sensitivity in $\theta_{13}$.

Similar plans are emerging worldwide, for instance in North America with the Deep Underground Science and
Engineering Laboratory (DUSEL) at Homestake (South Dakota), envisioning a deep underground facility coupled to 
US accelerator laboratories located at appropriate distances, for long baseline neutrino oscillation experiments~\cite{Kimyp}. In 
Asia, the Japanese High Energy Research Accelerator Research Organization (KEK) roadmap foresees extensions 
of the JPARC neutrino programme beyond the current T2K experiment by increasing the neutrino beam intensity 
and by constructing a new far detector in addition to SuperKamiokande~\cite{Hasegawayp}.

This international landscape underlines the ``global" nature of the project, with potential options being considered in 
several continents, jointly debated by the international scientific community, making all the more likely that only one such 
facility will be built worldwide.

\section{Conclusions and Outlook}

The LAGUNA consortium is studying the feasibility of a new large underground infrastructure in Europe able to host
next generation neutrino physics and astrophysics and proton decay experiments. Seven sites are presently being
considered. The study
aims at the comprehensive and coordinated technical assessment of each site, 
at a coherent cost estimation, and at a priority list for site selection, by Summer 2010. After that, a more focused
design study, aiming also at detector option selection, should be envisaged.
Future long baseline in Europe should consider a new beam line from CERN 
towards the chosen LAGUNA site.
The new large underground infrastructure could in the future be operated in connection with other, more advanced
neutrino beams like beta-beams or neutrino factories. 

\section*{Acknowledgements}

The LAGUNA design study is financed by FP7 Research Infrastructure "Design Studies", Grant Agreement No.
212343 FP7-INFRA-2007-1.

\end{document}